\documentclass[english]{article}
\usepackage[T1]{fontenc}
\usepackage[latin9]{inputenc}
\usepackage[a4paper]{geometry}
\usepackage{amsmath}
\usepackage{amssymb}

\makeatletter
\newcommand{\lyxaddress}[1]{
\par {\raggedright #1
\vspace{1.4em}
\noindent\par}
}

\makeatother

\usepackage{babel}
\begin{document}

\title{\textbf{Novel energy level structure of Dirac oscillator in magnetic
field }}

\author{Md. Moniruzzaman%
\thanks{Corresponding Author: Email: monir\_m17@yahoo.com, %
} %
\thanks{Permanent Address: Department of Physics, Mawlana Bhashani Science
and Technology University, Santosh, Tangail-1902, Bangladesh%
} and S. B. Faruque }

\maketitle

\lyxaddress{Department of Physics, Shahjalal University of Science and Technology,
Sylhet 3114, Bangladesh }
\begin{abstract}
We have presented an elegant high energy quantum problem, namely,
the full Dirac oscillator under axial magnetic field with its full
solution. We have found the energy spectrum which is rich and at the
same time has a novel structure. The quantized energy levels show
coupling of the oscillator frequency with the Larmor frequency in
the 2D surface where the electrons under consideration follow a 2D
oscillator. The axis in which magnetic field is pointed, the electrons
follow a 1D oscillator. There is also coupling between spin and orbital
motion and also a coupling between a resultant effect of orbital and
spin motion with Larmor precession. 

\end{abstract}

\section{Introduction}

Moshinsky and Szczepaniak {[}1{]} have introduced an interesting interaction
in the Dirac equation such that the Dirac Hamiltonian is linear in
both momentum and the coordinates. The resulting Dirac equation is
known as Dirac oscillator because it turns into a harmonic oscillator
with strong spin-orbit coupling in the non-relativistic limit. The
oscillator is achieved by introducing the coupling in the Dirac equation
as {[}1{]} 
\[
i\hbar\frac{\partial\varPsi}{\partial t}=\left[c\overrightarrow{\alpha}.\left(\overrightarrow{p}-im\beta\omega\overrightarrow{r}\right)+\beta mc^{2}\right]\varPsi,
\]
where $\varPsi$ is the four component bispinor corresponding to spin-half
particle, like electron, $c$ is the speed of light, $\overrightarrow{\alpha}$
and $\mathbf{\beta}$ are standard representation of Dirac matrices,
$\omega$ is the classical frequency of the oscillator. When $\omega=0$
, we recover the ordinary Dirac equation{[}2{]}.

The Dirac oscillator plays a vital role in the depiction of relativistic
many body problems {[}3, 4 and 5{]} and super symmetric relativistic
quantum mechanics {[}6, 7, and 8{]}. Dirac oscillator framework is
also proposed in quantum chromodynamics, mainly in connection with
quark confinement models in baryons and mesons {[}9{]}. Most thrilling
aspects of this oscillator are its association to quantum optics {[}10{]}.
It also maps into Anti-Jaynes-Cummings model for describing the interaction
of a atomic transition in two level systems {[}11, 12, and 13{]}.
Moreover, a lot of attention and many physical applications in various
branches of physics of this oscillator have been found, for example,
in semiconductor physics {[}14{]}, graphene {[}15{]} etc. Recently
an experimental demonstration of this oscillator has been initiated
{[}16{]}.

In this context, we consider (3+1) dimensional Dirac oscillator in
an external homogeneous magnetic field. We investigated mathematical
formulation of the oscillator problem and solved it. The quantized
energy spectrum for up and down spin is derived. In the non-relativistic
limit, this energy spectrum reduces to the spectrum of three dimensional
harmonic oscillator with a spin-orbit contribution, when the magnetic
field is absent. We present this problem and give a solution in section
2 and in section 3, we give a brief conclusion of the work.

\section{Dirac oscillator in magnetic field}

In presence of an external magnetic field, the (3+1) dimensional Dirac
oscillator is given by 

\begin{equation}
\left[c\overrightarrow{\alpha}.\left(\overrightarrow{p}+\frac{e}{c}\overrightarrow{A}-im\beta\omega\overrightarrow{r}\right)+\beta mc^{2}\right]\psi=E\psi,
\end{equation}
where $\overrightarrow{A}$ is vector potential and $-e(e>0)$ is
the charge of the oscillator which is considered to be an electron
here. We consider the uniform magnetic field to be along the $z$-
direction and vector potential to be in the symmetric gauge as $\overrightarrow{A}=\left(-\frac{B}{2}y,\frac{B}{2}x,0\right)$,
where $B$ is the magnetic field strength. Using the following standard
form of $\overrightarrow{\alpha}$ and $\beta$:

\[
\overrightarrow{\alpha}=\left(\begin{array}{cc}
0 & \overrightarrow{\sigma}\\
\overrightarrow{\sigma} & 0
\end{array}\right)
\]
and 

\begin{equation}
\beta=\left(\begin{array}{cc}
I & 0\\
0 & -I
\end{array}\right)
\end{equation}
and two-component form for $\psi$ as

\begin{equation}
\psi=\left(\begin{array}{c}
\psi_{a}\\
\psi_{b}
\end{array}\right),
\end{equation}
we obtain the following two simultaneous coupled equations:

\begin{equation}
\left(c\overrightarrow{\sigma}.\overrightarrow{p}+e\overrightarrow{\sigma}.\overrightarrow{A}\right)\psi_{b}+imc\omega\overrightarrow{\sigma}.\overrightarrow{r}\psi_{b}=\left(E-mc^{2}\right)\psi_{a}
\end{equation}
and
\begin{equation}
\left(c\overrightarrow{\sigma}.\overrightarrow{p}+e\overrightarrow{\sigma}.\overrightarrow{A}\right)\psi_{a}-imc\omega\overrightarrow{\sigma}.\overrightarrow{r}\psi_{a}=\left(E+mc^{2}\right)\psi_{b}.
\end{equation}
The system gives the following two equations 
\[
[c^{2}\left(\overrightarrow{\sigma}.\overrightarrow{p}\right)^{2}+ce\left(\overrightarrow{\sigma}.\overrightarrow{A}\right)\left(\overrightarrow{\sigma}.\overrightarrow{p}\right)+ce\left(\overrightarrow{\sigma}.\overrightarrow{p}\right)\left(\overrightarrow{\sigma}.\overrightarrow{A}\right)+e^{2}\left(\overrightarrow{\sigma}.\overrightarrow{A}\right)^{2}+imc^{2}\omega\left[\left(\overrightarrow{\sigma}.\overrightarrow{r}\right),\left(\overrightarrow{\sigma}.\overrightarrow{p}\right)\right]
\]
\begin{equation}
+imce\omega\left[\left(\overrightarrow{\sigma}.\overrightarrow{r}\right),\left(\overrightarrow{\sigma}.\overrightarrow{A}\right)\right]+m^{2}c^{2}\omega^{2}\left(\overrightarrow{\sigma}.\overrightarrow{r}\right)^{2}]\psi_{a}=\left(E^{2}-m^{2}c^{4}\right)\psi_{a}
\end{equation}
and
\[
[c^{2}\left(\overrightarrow{\sigma}.\overrightarrow{p}\right)^{2}+ce\left(\overrightarrow{\sigma}.\overrightarrow{A}\right)\left(\overrightarrow{\sigma}.\overrightarrow{p}\right)+ce\left(\overrightarrow{\sigma}.\overrightarrow{p}\right)\left(\overrightarrow{\sigma}.\overrightarrow{A}\right)+e^{2}\left(\overrightarrow{\sigma}.\overrightarrow{A}\right)^{2}-imc^{2}\omega\left[\left(\overrightarrow{\sigma}.\overrightarrow{r}\right),\left(\overrightarrow{\sigma}.\overrightarrow{p}\right)\right]
\]
\begin{equation}
-imce\omega\left[\left(\overrightarrow{\sigma}.\overrightarrow{r}\right),\left(\overrightarrow{\sigma}.\overrightarrow{A}\right)\right]+m^{2}c^{2}\omega^{2}\left(\overrightarrow{\sigma}.\overrightarrow{r}\right)^{2}]\psi_{b}=\left(E^{2}-m^{2}c^{4}\right)\psi_{b}.
\end{equation}
One can easily show that 

\begin{equation}
\left(\overrightarrow{\sigma}.\overrightarrow{p}\right)^{2}=p^{2},
\end{equation}
\begin{equation}
\left(\overrightarrow{\sigma}.\overrightarrow{A}\right)\left(\overrightarrow{\sigma}.\overrightarrow{p}\right)+\left(\overrightarrow{\sigma}.\overrightarrow{p}\right)\left(\overrightarrow{\sigma}.\overrightarrow{A}\right)=BL_{z}+B\hbar\sigma_{z},
\end{equation}
\begin{equation}
\left(\overrightarrow{\sigma}.\overrightarrow{A}\right)^{2}=A^{2},
\end{equation}

\begin{equation}
\left[\left(\overrightarrow{\sigma}.\overrightarrow{r}\right),\left(\overrightarrow{\sigma}.\overrightarrow{p}\right)\right]=\frac{4i}{\hbar}\overrightarrow{L}.\overrightarrow{S}+3i\hbar,
\end{equation}
where electron spin $\overrightarrow{S}=\frac{\hbar}{2}\overrightarrow{\sigma}$,
\begin{equation}
\left[\left(\overrightarrow{\sigma}.\overrightarrow{r}\right),\left(\overrightarrow{\sigma}.\overrightarrow{A}\right)\right]=ir^{2}[\overrightarrow{\sigma}.\overrightarrow{B}-\left(\overrightarrow{\sigma}.\hat{r}\right)\left(\overrightarrow{B}.\hat{r}\right)]
\end{equation}
and
\begin{equation}
\left(\overrightarrow{\sigma}.\overrightarrow{r}\right)^{2}=r^{2}.
\end{equation}
Using these, Eqs. (6) and (7) can be cast into 
\[
\left[c^{2}p^{2}+m^{2}c^{2}\omega^{2}r^{2}+ceB(L_{Z}+\hbar\sigma_{z})+e^{2}A^{2}-\frac{4mc^{2}\omega}{\hbar}\overrightarrow{L}.\overrightarrow{S}-mce\omega r^{2}[\sigma_{z}B-\left(\overrightarrow{\sigma}.\hat{r}\right)\left(\overrightarrow{B}.\hat{r}\right)]\right]\psi_{a}
\]
\begin{equation}
=\left(E^{2}-m^{2}c^{4}+3mc^{2}\hbar\omega\right)\psi_{a}
\end{equation}
and
\[
\left[c^{2}p^{2}+m^{2}c^{2}\omega^{2}r^{2}+ceB(L_{Z}+\hbar\sigma_{z})+e^{2}A^{2}+\frac{4mc^{2}\omega}{\hbar}\overrightarrow{L}.\overrightarrow{S}+mce\omega r^{2}[\sigma_{z}B-\left(\overrightarrow{\sigma}.\hat{r}\right)\left(\overrightarrow{B}.\hat{r}\right)]\right]\psi_{b}
\]
\begin{equation}
=\left(E^{2}-m^{2}c^{4}-3mc^{2}\hbar\omega\right)\psi_{b}.
\end{equation}
Since $\overrightarrow{B}$ is along the z-direction, and in magnetic
field spin of an electron precesses around $\overrightarrow{B}$,
the effective component of $\overrightarrow{\sigma}$ is $\sigma_{z}$.
Hence, the second term within the square bracket in the 6th term of
the both equations should reduce as follows: 
\[
\left(\overrightarrow{\sigma}.\hat{r}\right)\left(\overrightarrow{B}.\hat{r}\right)=\sigma_{z}B\cos^{2}\theta,
\]
where $\theta$ is the azimuthal angle. 

Then we have
\[
\left[c^{2}p^{2}+m^{2}c^{2}\omega^{2}r^{2}+ceB(L_{Z}+\hbar\sigma_{z})+\frac{e^{2}B^{2}}{4}r^{2}\sin^{2}\theta-\frac{4mc^{2}\omega}{\hbar}\overrightarrow{L}.\overrightarrow{S}-mce\omega\sigma_{z}Br^{2}\sin^{2}\theta\right]\psi_{a}
\]
\begin{equation}
=\left(E^{2}-m^{2}c^{4}+3mc^{2}\hbar\omega\right)\psi_{a}
\end{equation}
and
\[
\left[c^{2}p^{2}+m^{2}c^{2}\omega^{2}r^{2}+ceB(L_{Z}+\hbar\sigma_{z})+\frac{e^{2}B^{2}}{4}r^{2}\sin^{2}\theta+\frac{4mc^{2}\omega}{\hbar}\overrightarrow{L}.\overrightarrow{S}+mce\omega\sigma_{z}Br^{2}\sin^{2}\theta\right]\psi_{b}
\]
\begin{equation}
=\left(E^{2}-m^{2}c^{4}-3mc^{2}\hbar\omega\right)\psi_{b}.
\end{equation}
The first two terms of the left hand side of Eqs.(16) and (17) constitute
Hamiltonian of 3 dinensoinal harmonic oscillator. Third term reveals
an interaction between $z$ component of total magnetic moment and
the magnetic field and the fourth term reveals interaction between
electric charge and the magnetic field. The fifth term appears due
to spin-orbit interaction and sixth term due to an interaction between
electron spin and magnetic field. 

We can find closed form solution of Eqs.(16) and (17) using either
$\mid Y_{l}^{m_{l}}(\theta,\phi)\chi_{m_{s}}>$ basis or $\mid e^{im_{l}\phi}\chi_{m_{s}}>$
basis, where the quantum number $m_{l}$ is associated with the eigenfunctions
$e^{im_{l}\phi}$ of $L_{z}$ and $S_{z}\chi_{m_{s}}=m_{s}\hbar\chi_{m_{s}}$,
$m_{s}=\pm\frac{1}{2}$. In both of those bases, the eigen value of
$\overrightarrow{L}.\overrightarrow{S}$ is $\hbar^{2}m_{l}m_{s}$.
We prefer now to work with the $\mid e^{im_{l}\phi}\chi_{m_{s}}>$
basis.

Since the system possesses cylindrical symmetry, in cylindrical coordinates
$\rho$, $\phi$ and $z$ , where $\rho^{2}=x^{2}+y^{2}$, Eqs. (16)
and (17) can be written respectively as
\[
[-c^{2}\hbar^{2}\left\{ \frac{1}{\rho}\frac{\partial}{\partial\rho}\left(\rho\frac{\partial}{\partial\rho}\right)+\frac{1}{\rho^{2}}\frac{\partial^{2}}{\partial\phi^{2}}+\frac{\partial^{2}}{\partial z^{2}}\right\} +m^{2}c^{2}\omega^{2}\rho^{2}+m^{2}c^{2}\omega^{2}z^{2}+ceB(L_{Z}+\hbar\sigma_{z})+\frac{e^{2}B^{2}}{4}\rho^{2}-\frac{4mc^{2}\omega}{\hbar}\overrightarrow{L}.\overrightarrow{S}
\]
\begin{equation}
-mce\omega\sigma_{z}B\rho^{2}]\psi_{a}=\left(E^{2}-m^{2}c^{4}+3mc^{2}\hbar\omega\right)\psi_{a}
\end{equation}
and
\[
[-c^{2}\hbar^{2}\left\{ \frac{1}{\rho}\frac{\partial}{\partial\rho}\left(\rho\frac{\partial}{\partial\rho}\right)+\frac{1}{\rho^{2}}\frac{\partial^{2}}{\partial\phi^{2}}+\frac{\partial^{2}}{\partial z^{2}}\right\} +m^{2}c^{2}\omega^{2}\rho^{2}+m^{2}c^{2}\omega^{2}z^{2}+ceB(L_{Z}+\hbar\sigma_{z})+\frac{e^{2}B^{2}}{4}\rho^{2}+\frac{4mc^{2}\omega}{\hbar}\overrightarrow{L}.\overrightarrow{S}
\]

\begin{equation}
+mce\omega\sigma_{z}B\rho^{2}]\psi_{b}=\left(E^{2}-m^{2}c^{4}-3mc^{2}\hbar\omega\right)\psi_{b}.
\end{equation}
As the equations involve partial derivatives of $\rho$ and $z$ and
operators of z-component of angular momentum and spin, we take the
solutions in the form : 
\begin{equation}
\psi_{a}=F_{a}(\rho)G_{a}(z)e^{im_{l}\phi}\chi_{m_{s}=-\frac{1}{2}}
\end{equation}
and

\begin{equation}
\psi_{b}=F_{b}(\rho)G_{b}(z)e^{im_{l}\phi}\chi_{m_{s}=\frac{1}{2}}.
\end{equation}
 Then we get from Eq.(18) 
\[
-\frac{1}{F_{a}(\rho)}c^{2}\hbar^{2}\frac{1}{\rho}\frac{\partial}{\partial\rho}\left(\rho\frac{\partial F_{a}(\rho)}{\partial\rho}\right)+c^{2}\hbar^{2}\frac{m_{l}^{2}}{\rho^{2}}+\left(\frac{B^{2}e^{2}}{4}+mce\omega B+m^{2}c^{2}\omega^{2}\right)\rho^{2}-c^{2}\hbar^{2}\frac{1}{G_{a}(z)}\frac{\partial^{2}G_{a}(z)}{\partial z^{2}}+m^{2}c^{2}\omega^{2}z^{2}
\]
\begin{equation}
=E^{2}-m^{2}c^{4}+3mc^{2}\hbar\omega-ceB(m_{l}-1)\hbar-2mc^{2}\omega m_{l}\hbar.
\end{equation}

Let
\begin{equation}
E^{2}-m^{2}c^{4}+3mc^{2}\hbar\omega-ceB(m_{l}-1)\hbar-2mc^{2}\omega m_{l}\hbar=\lambda,
\end{equation}
 where $\lambda$ is a constant. Hence, the Eq.(22) demands 
\begin{equation}
-c^{2}\hbar^{2}\frac{1}{G_{a}(z)}\frac{d^{2}G_{a}(z)}{dz^{2}}+m^{2}c^{2}\omega^{2}z^{2}=\epsilon,
\end{equation}
where $\epsilon$ is another constant. Then we have 
\begin{equation}
-c^{2}\hbar^{2}\frac{1}{\rho}\frac{d}{d\rho}\left(\rho\frac{dF_{a}(\rho)}{d\rho}\right)+c^{2}\hbar^{2}\frac{m_{l}^{2}}{\rho^{2}}F_{a}(\rho)+\left(\frac{B^{2}e^{2}}{4}+mce\omega B+m^{2}c^{2}\omega^{2}\right)\rho^{2}F_{a}(\rho)+(\epsilon-\lambda)F_{a}(\rho)=0.
\end{equation}
Multiplying Eq.(24) by $\frac{1}{2mc^{2}}$, we have 
\begin{equation}
-\frac{\hbar^{2}}{2m}\frac{d^{2}G_{a}(z)}{dz^{2}}+\frac{1}{2}m\omega^{2}z^{2}G_{a}(z)=\frac{\epsilon}{2mc^{2}}G_{a}(z)
\end{equation}
This is the simple harmonic oscillator Schrogringer equation and hence,
\begin{equation}
\epsilon=2\left(n+\frac{1}{2}\right)\hbar\omega mc^{2},n=0,1,2...
\end{equation}
and $G_{a}(z)$ is the wave function of simple harmonic oscillator.

Eq.(25) can be written as 
\begin{equation}
\frac{d^{2}F_{a}(\rho)}{d\rho^{2}}+\frac{1}{\rho}\frac{dF_{a}(\rho)}{d\rho}-\frac{m_{l}^{2}}{\rho^{2}}F_{a}(\rho)-\left(\frac{B^{2}e^{2}}{4c^{2}\hbar^{2}}+\frac{me\omega B}{\hbar^{2}c}+\frac{m^{2}\omega^{2}}{\hbar^{2}}\right)\rho^{2}F_{a}(\rho)+\frac{1}{c^{2}\hbar^{2}}(\lambda-\epsilon)F_{a}(\rho)=0
\end{equation}
Now, performing a change of variable 

\begin{equation}
\xi=\sqrt[4]{\frac{B^{2}e^{2}}{4c^{2}\hbar^{2}}+\frac{me\omega B}{\hbar^{2}c}+\frac{m^{2}\omega^{2}}{\hbar^{2}}}\varrho,
\end{equation}
we obtain from Eq.(28)
\begin{equation}
\frac{d^{2}F_{a}(\xi)}{d\xi^{2}}+\frac{1}{\xi}\frac{dF_{a}(\xi)}{d\xi}-\frac{m_{l}^{2}}{\xi^{2}}F_{a}(\xi)-\xi^{2}F_{a}(\xi)+DF_{a}(\xi)=0,
\end{equation}
where
\begin{equation}
D=\frac{\lambda-\epsilon}{c^{2}\hbar^{2}\sqrt{\frac{B^{2}e^{2}}{4c^{2}\hbar^{2}}+\frac{me\omega B}{\hbar^{2}c}+\frac{m^{2}\omega^{2}}{\hbar^{2}}}}.
\end{equation}
According to {[}17{]}, the corresponding solutions are 
\begin{equation}
D=2(N+1),\begin{array}{c}
\end{array}N=0,1,2...
\end{equation}
and 
\begin{equation}
F_{a}(\xi)=Ae^{-\frac{1}{2}\xi^{2}}f_{m_{l}+k}^{m_{l}}(\xi),
\end{equation}
where $A$ is a normalization constant and $f_{m_{l}+k}^{m_{l}}(\xi)$
satisfy the equation
\begin{equation}
\xi\frac{d^{2}f_{m_{l}+k}^{m_{l}}(\xi)}{d\xi^{2}}+\left(m_{l}+1-\xi\right)\frac{df_{m_{l}+k}^{m_{l}}(\xi)}{d\xi}+kf_{m_{l}+k}(\xi)=0,
\end{equation}
\begin{equation}
k=\frac{1}{2}\left(N-m_{l}\right)=\begin{cases}
0,1,2.........\frac{N}{2}, & \begin{array}{cccccc}
for & N & is & an & even & integer\end{array}\\
0,1,2.........\frac{N-1}{2}, & \begin{array}{cccccc}
for & N & is & an & odd & intiger\end{array}
\end{cases},
\end{equation}
 and $m_{l}$ can take $\left(N/2\right)+1$ values for an even intiger
$N$ and $N+1$ values for an odd intiger $N$.

Equation (32) gives the energy spectrum of the (3+1) dimensional Dirac
oscillator in presence of magnetic field:
\begin{equation}
E^{2}-m^{2}c^{4}=2mc^{2}\left[\left(N+1\right)\hbar\left(\omega+\omega_{L}\right)+(n+\frac{1}{2})\hbar\omega+\left(m_{l}-\frac{3}{2}\right)\hbar\omega+(m_{l}-1)\hbar\omega_{L}\right]\begin{array}{ccccc}
 & for & m_{s}=-\frac{1}{2}\end{array},
\end{equation}
where the Larmor frequency $\omega_{L}$ is 
\begin{equation}
\omega_{L}=\frac{eB}{2mc}.
\end{equation}
The Larmor frequency occurs in the energy spectrum due to interaction
between electric charge and the magnetic field. The third term in
the square bracket of Eq.(36) appears for spin-orbit coupling and
the fourth term for an interaction between total magnetic moment and
the magnetic field.

Now, the other component of $\psi$ is 
\begin{equation}
\psi_{b}=F_{b}(\rho)G_{b}(z)e^{im_{l}\phi}\chi_{m_{s}=\frac{1}{2}}=F_{b}(\xi)G_{b}(z)e^{im_{l}\phi}\chi_{m_{s}=\frac{1}{2}}=A^{\prime}e^{-\frac{1}{2}\xi^{2}}f_{m_{l}+k}^{m_{l}}(\xi)G_{b}(z)e^{im_{l}\phi}\chi_{m_{s}=\frac{1}{2}}
\end{equation}
and the corresponding energy spectrum is 
\begin{equation}
E^{2}-m^{2}c^{4}=2mc^{2}\left[\left(N+1\right)\hbar\left(\omega+\omega_{L}\right)+(n+\frac{1}{2})\hbar\omega+\left(m_{l}+\frac{3}{2}\right)\hbar\omega+(m_{l}+1)\hbar\omega_{L}\right]\begin{array}{ccccc}
 & for & m_{s}=\frac{1}{2}\end{array}.
\end{equation}

For $B=0$
\[
E^{2}-m^{2}c^{4}=2mc^{2}\left[\left(N+n+\frac{3}{2}\right)\hbar\omega+\left(m_{l}-\frac{3}{2}\right)\hbar\omega\right]
\]
\begin{equation}
=2mc^{2}\left[\left(n^{\prime}+\frac{3}{2}\right)\hbar\omega+\left(m_{l}-\frac{3}{2}\right)\hbar\omega\right]\begin{array}{ccccc}
 & for & m_{s}=-\frac{1}{2}\end{array}
\end{equation}
and
\[
E^{2}-m^{2}c^{4}=2mc^{2}\left[\left(N+n+\frac{3}{2}\right)\hbar\omega+\left(m_{l}+\frac{3}{2}\right)\hbar\omega\right]
\]
\begin{equation}
=2mc^{2}\left[\left(n^{\prime}+\frac{3}{2}\right)\hbar\omega+\left(m_{l}+\frac{3}{2}\right)\hbar\omega\right]\begin{array}{ccccc}
 & for & m_{s}=\frac{1}{2}\end{array},
\end{equation}
where 
\begin{equation}
n^{\prime}=N+n=0,1,2....
\end{equation}
These (Eqs.(40) and (41)) are of the same nature as in spherically
symmetric Dirac oscillator {[}18{]}; the difference occurred only
because of the cylindrical symmetry we have assumed from the very
beginning of the solution procedure. 

The non relativistic limit of the energy spectrum is obtained by setting
$E=mc^{2}+K$ with the consideration $K\ll mc^{2}$ and we get 
\begin{equation}
K=\left(N+1\right)\hbar\left(\omega+\omega_{L}\right)+(n+\frac{1}{2})\hbar\omega+\left(m_{l}-\frac{3}{2}\right)\hbar\omega+(m_{l}-1)\hbar\omega_{L}\begin{array}{ccccc}
 & for & m_{s}=-\frac{1}{2}\end{array}
\end{equation}
and
\begin{equation}
K=\left(N+1\right)\hbar\left(\omega+\omega_{L}\right)+(n+\frac{1}{2})\hbar\omega+\left(m_{l}+\frac{3}{2}\right)\hbar\omega+(m_{l}+1)\hbar\omega_{L}\begin{array}{ccccc}
 & for & m_{s}=\frac{1}{2}.\end{array}
\end{equation}
For $B=0$ 
\begin{equation}
K=\left(n^{\prime}+\frac{3}{2}\right)\hbar\omega+\left(m_{l}-\frac{3}{2}\right)\hbar\omega\begin{array}{ccccc}
 & for & m_{s}=-\frac{1}{2}\end{array}
\end{equation}
and
\begin{equation}
K=\left(n^{\prime}+\frac{3}{2}\right)\hbar\omega+\left(m_{l}+\frac{3}{2}\right)\hbar\omega\begin{array}{ccccc}
 & for & m_{s}=\frac{1}{2}.\end{array}
\end{equation}
Except spin-orbit contribution terms $\left(m_{l}-\frac{3}{2}\right)\hbar\omega$
in Eq.(45) and $\left(m_{l}+\frac{3}{2}\right)\hbar\omega$ in Eq.(46),
these are exactly the energy spectrum of 3D non-relativistic harmonic
oscillator.

If we take the solutions in the form : 
\begin{equation}
\psi_{a}=F_{a}(\rho)G_{a}(z)e^{im_{l}\phi}\chi_{m_{s}=\frac{1}{2}}
\end{equation}
and

\begin{equation}
\psi_{b}=F_{b}(\rho)G_{b}(z)e^{im_{_{l}}\phi}\chi_{m_{s}=-\frac{1}{2}},
\end{equation}
the corresponding energy spectrum of $\psi_{a}$ and $\psi_{b}$ are
given by respectively as 
\begin{equation}
E^{2}-m^{2}c^{4}=2mc^{2}\left[\left(N+1\right)\hbar\left(\omega-\omega_{L}\right)+(n+\frac{1}{2})\hbar\omega-\left(m_{l}+\frac{3}{2}\right)\hbar\omega+(m_{l}+1)\hbar\omega_{L}\right]\begin{array}{ccccc}
 & for & m_{s}=\frac{1}{2}\end{array}
\end{equation}
and
\begin{equation}
E^{2}-m^{2}c^{4}=2mc^{2}\left[\left(N+1\right)\hbar\left(\omega-\omega_{L}\right)+(n+\frac{1}{2})\hbar\omega-\left(m_{l}-\frac{3}{2}\right)\hbar\omega+(m_{l}-1)\hbar\omega_{L}\right]\begin{array}{ccccc}
 & for & m_{s}=-\frac{1}{2}.\end{array}
\end{equation}

\section{Conclusion }

We have solved the (3+1) Dirac oscillator with magnetic field and
found the energy spectrum. The system is cylindrically symmetric:
there are basically two oscillators, one is a 2D oscillator in the
$x-y$ plane and the other is a 1D oscillator in the $z$- direction.
The energy spectrum shows this splitting exactly. Moreover, there
appears an oscillation which is basically a Larmor precession, coupled
(i) with the 2D oscillator in $x-y$ plane and (ii) coupled with a
net effect of orbital and spin angular momentum. The 2nd coupling
is shown in the last term in Eqs.(36) and (39) or (49) and (50). There
is also spin-orbit coupling coupled with the Dirac oscillator in the
third term in Eqs.(36) and (39) or (49) and (50). The non-relativistic
limit of the energy spectrum is rightly that of 3D harmonic oscillators
plus spin-orbit coupling. Everything appeared smoothly in our calculation
because of the initial decomposition of the problem by a cylindrical
symmetry. In the absence of magnetic field, the problem shows characteristics
of a 3D Dirac oscillator as can be verified with what is available
in literature. The energy spectrum is novel and rich. This oscillator
might find application in quantum optics. The system can easily be
manipulated using photon beams or electromagnetic signal and made
to absorb and emit radiation with energy $\hbar\omega$ or $\hbar\omega_{L}$
or $\hbar\left(\omega+\omega_{L}\right)$ or $\hbar\left(\omega-\omega_{L}\right)$.
This is clear from the expression of the energy levels. 

In conclusion, a very attractive quantum system has been elucidated
which is a 3D Dirac oscillator immersed in axial magnetic field.

\section*{Acknowledgment}

We are grateful to the reviewer whose constructive suggestions were
greatly useful in correcting our calculation and in improving the
presentation of the work.

\end{document}